# The evolution of cooperation: an evolutionary advantage of individuals impedes the evolution of the population

Xiaoliang Wang[1,3]*, Andrew Harrison[2]*

1 College of Life Sciences, Zhejiang University, Hangzhou 310058, China

2 Department of Mathematical Sciences, University of Essex, Colchester CO4 3SQ, UK

3 School of Physical Sciences, University of Science and Technology of China, Hefei 230026, China

*Correspondence: (X.W.) wxliang@mail.ustc.edu.cn; (A.H.) harry@essex.ac.uk;

**Abstract**

Range expansion is a universal process in biological systems, and therefore plays a part in biological evolution. Using a quantitative individual-based method based on the stochastic process, we identify that enhancing the inherent self-proliferation advantage of cooperators relative to defectors is a more effective channel to promote the evolution of cooperation in range expansion than weakening the benefit acquisition of defectors from cooperators. With this self-proliferation advantage, cooperators can rapidly colonize virgin space and establish spatial segregation more readily, which acts like a protective shield to further promote the evolution of cooperation in return. We also show that lower cell density and migration rate have a positive effect on the competition of cooperators with defectors.

Biological evolution is based on competition between individuals and should therefore favor selfish behaviors. However, we observe a counterintuitive phenomenon that the evolution of a population is impeded by the fitness-enhancing chemotactic movement of individuals. This highlights a conflict between the interests of the individual and the population. The short-sighted selfish behavior of individuals may not be that favored in the competition between populations. Such information provides important implications for the handling of cooperation.

## 1. Introduction

Biological evolution is closely related to self-organized biological pattern formation. In evolution, through cooperation among their components, complex organizations exhibit emergent behaviors (Fig. 1). This has happened in biological systems at all levels [1,2]. For example, genes cooperate in a genome, organelles cooperate in a cell and organisms cooperate in an ecological community. However, it's not easy to understand the maintenance of cooperation in biological evolution, since organisms with cooperative genes often have no competitive advantage relative to those with defective/cheating genes whereas evolution is based on competition between individuals [3] (in Prisoner's dilemma, defection is often the evolutionarily stable strategy, ESS [4]).

Explaining the evolution of cooperation has long been of interest and has not yet been fully understood. Many theories and mechanisms have been proposed, including kin selection [3,5,6], reciprocity [7-9], group selection [10-13], policing and punishment [14,15], and biological range expansion [16-19]. Kin selection is the most famous, which argues that the benefits produced by cooperation can be directed to those individuals with inclusive genes (i.e. relatives), and therefore facilitate the maintenance of cooperative behaviors. Such explanations strengthen Darwin natural selection theory, which is usually considered solely on individual behaviors.

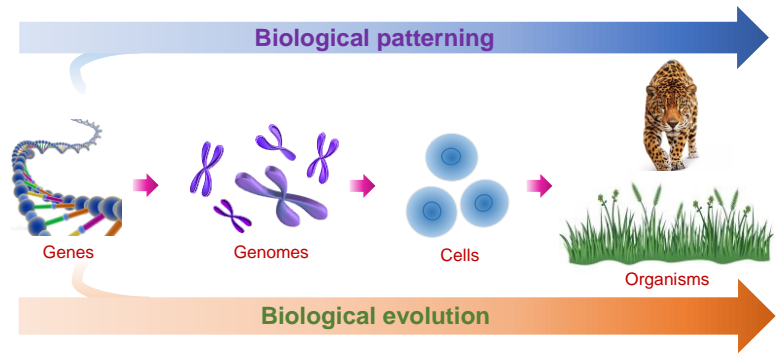

**Figure 1.** Schematic diagram of connection between biological pattern formation and biological evolution.

In recent years, evolutionary biology is studied quantitatively in both theory and experiments. Especially, quantitative agent-based models [20-22] have been developed for finite populations, which usually exhibit prominent random effects. These models have greatly improved our understanding of biological systems. In this article, we utilize the quantitative approach and model microorganisms to address a key question of biological evolution: how can cooperation be maintained in the presence of cheating? Real biological systems tend to expand their range by moving into nearby free space [20,23-27], resulting from the inherent growth and reproduction of individuals. Here we will explore the problem of biological systems with range expansion. Our ultimate intention here is to provide new insights into the evolution of cooperation.

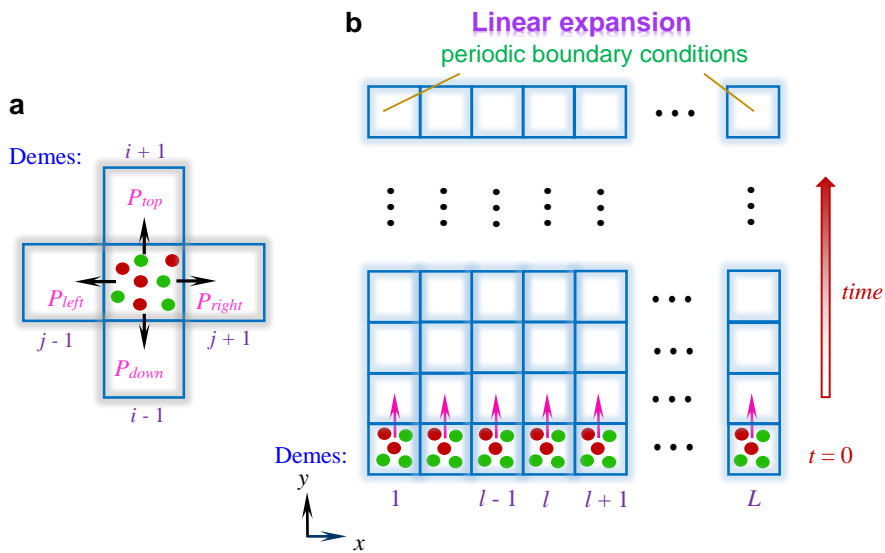

**Figure 2.** Illustration of (**a**) 2D on-lattice simulation and (**b**) linear expansion of a population. One line of demes equates about one generation.

## 2. Models and methods

We introduce here a two-dimensional on-lattice simulation, which is based on stochastic processes, to deal with biological evolutionary dynamics in range expansion. In this model, the number of individuals (i.e. the population size) in demes located on spatial lattices is not necessary to be kept invariant at every time step and the motions of individuals are not limited within the fixed space, either. As shown in Fig. 2a, during each time step, one individual in each deme has a probability to reproduce *in situ* and a probability to migrate into one of the four nearest demes. We take linear expansion as an example to show the general evolutionary process (Fig. 2b). Initially, L demes containing well-mixed populations of the same size are arranged on a line. The next generation is produced by allowing the current generation to reproduce one by one until the number of each deme reaches the maximum population size *N* (cell density). Before the production of next generation is finished, at each time step individuals are allowed to exchange positions with its neighbors from the nearest demes. Each step happens randomly. To capture the general feature of a biological system, periodic boundary conditions are used for both right and left boundaries.

With the development of biotechnology, especially the synthetic biology, researchers recently start to use model microorganisms (like *E. coli* and *yeast*) to test the evolutionary theory mainly due to their short generation time and simple life system (convenient for genetic manipulation) [15]. In the following, the 2D on-lattice simulation is further introduced to deal with evolutionary games and chemotaxis of microorganisms.

## 2.1 Evolutionary games of microorganisms

In two-player games, a 2×2 payoff matrix is often used to describe the benefit distribution for the strategies A and B conducted between players (Table 1). In evolutionary biology, strategies A and B usually denote the cooperation and defection, respectively. Explanations for elements a, b, c, and d are as follows: when both players select cooperation, the benefits they both get are a; when one side conducts cooperation and the other selects defection, the side which selects cooperation gets the benefit of b, while the benefit the side which conducts defection gets is c; when both sides conduct defection, the benefits they both get are d. In games between microorganisms, a, b, c and d all correspond to the fitness (i.e. growth rate). The difference is that a and d are the inherent growth rates of species A and B (e.g. the standard growth rate of *E. coli* strains is about 1 h$^{-1}$ [28]), while growth rates b and c only arise from social interactions (i.e. competition) between these two species.

Table 1: Benefits distribution in A-B strategy games.

|   | A | B |
|---|---|---|
| A | a | b |
| B | c | d |

We are mainly concerned by how cooperation can evolve in Prisoner's dilemma, i.e. a < c and b < d. In such a situation, cooperators usually have higher self-reproduction rates than defectors when they don't meet (i.e. a > d). When cooperators and defectors meet, however, defectors will cause damage to cooperators through social interactions between them, leading the later to have lower fitness since defectors often gain benefits from the former without paying the cost. In this study, we are mainly concerned by such a situation (Prisoner's dilemma).

- Reproduction probability of microorganisms

In random 2D on-lattice model, the growth rate of organisms is replaced by the reproduction probability. For microorganisms, an apparent feature is that the growth rate of them has a strong dependence on the population density. The growth rate is approximately proportional to population density when the density is low (this phenomenon is called the Allee effect [29]), and it has a maximum at intermediate cell densities [30]. Considering this fact, the reproduction probabilities per unit time for both species A and B in each deme are taken to be proportional to the fraction of cells in the deme and the fraction of vacancies in that deme [21]:

$$P_{A(i,j)} = \left[af_{i,j} + b(1-f_{i,j})\right]\frac{N_{t(i,j)}(N - N_{t(i,j)})}{N^2} \quad (1)$$

$$P_{B(i,j)} = \left[cf_{i,j} + d(1-f_{i,j})\right]\frac{N_{t(i,j)}(N - N_{t(i,j)})}{N^2} \quad (2)$$

where (i,j) denotes the position of demes, $N_t$ is the total number of individuals in demes and $f$ is frequency of species A in local demes.

- Migration probability of microorganisms

For the dense packing of cells, there is an obvious crowding effect. In such a case, the migration ability (motility) of individuals will be influenced by the cell density. In line with Korolev et al. [21], the probability that an individual from one deme migrates into a nearest neighbor deme in unit time is taken to be proportional to the fraction of cells in the deme and the fraction of vacancies in that neighbor deme. For species A and B, the probabilities per unit time migrating to the four nearest neighbor demes (Fig. 2a) are expressed as:

$$P_{top} = m_{A,B}\frac{N_{t(i,j)}(N - N_{t(i+1,j)})}{N^2} \quad (3)$$

$$P_{down} = m_{A,B}\frac{N_{t(i,j)}(N - N_{t(i-1,j)})}{N^2} \quad (4)$$

$$P_{right} = m_{A,B}\frac{N_{t(i,j)}(N - N_{t(i,j+1)})}{N^2} \quad (5)$$

$$P_{left} = m_{A,B}\frac{N_{t(i,j)}(N - N_{t(i,j-1)})}{N^2} \quad (6)$$

where $m$ is the inherent migration probability per unit time of individuals.

## 2.2 Chemotactic movement

After evolving for a long-term evolution, many smart species including microorganisms become equipped with chemotaxis behavior that leads to enhanced survival. The chemotactic motion which is guided by chemical cues actually means the anisotropic selection of motion directions. In simulations, for simplicity the chemotactic motion of individuals can be carried out by letting them select the motion direction with the probability proportional to positive gradient of cooperators (or the negative gradient of defectors). In case of species A, namely cooperators, we can let them partially move to the position with less defectors (species B) through the following method:

Partial amounts of direction selecting:

$$a_1 = (N - N_{t(i+1,j)}) + k_A(N - N_{Bt(i+1,j)}) \quad (7)$$
$$a_2 = (N - N_{t(i-1,j)}) + k_A(N - N_{Bt(i-1,j)}) \quad (8)$$
$$a_3 = (N - N_{t(i,j+1)}) + k_A(N - N_{Bt(i,j+1)}) \quad (9)$$
$$a_4 = (N - N_{t(i,j-1)}) + k_A(N - N_{Bt(i,j-1)}) \quad (10)$$

Selection probability of motion directions:

$$P_{i+1,j} = \frac{a_1}{a_1 + a_2 + a_3 + a_4} \quad (11)$$

$$P_{i-1,j} = \frac{a_2}{a_1 + a_2 + a_3 + a_4} \quad (12)$$

$$P_{i,j+1} = \frac{a_3}{a_1 + a_2 + a_3 + a_4} \quad (13)$$

$$P_{i,j-1} = \frac{a_4}{a_1 + a_2 + a_3 + a_4} \quad (14)$$

Similarly, for species B (defectors), the chemotactic motion can be realized through the following:

Partial amounts of direction selecting:

$$a_1 = (N - N_{t(i+1,j)}) + k_B N_{At(i+1,j)} \quad (15)$$
$$a_2 = (N - N_{t(i-1,j)}) + k_B N_{At(i-1,j)} \quad (16)$$
$$a_3 = (N - N_{t(i,j+1)}) + k_B N_{At(i,j+1)} \quad (17)$$
$$a_4 = (N - N_{t(i,j-1)}) + k_B N_{At(i,j-1)} \quad (18)$$

Selection probability of motion directions:

$$P_{i+1,j} = \frac{a_1}{a_1 + a_2 + a_3 + a_4} \quad (19)$$

$$P_{i-1,j} = \frac{a_2}{a_1 + a_2 + a_3 + a_4} \quad (20)$$

$$P_{i,j+1} = \frac{a_3}{a_1 + a_2 + a_3 + a_4} \quad (21)$$

$$P_{i,j-1} = \frac{a_4}{a_1 + a_2 + a_3 + a_4} \quad (22)$$

where $k_A$ and $k_B$ are the parameter which can be used to tune the chemotaxis strength and $N_{At}$ and $N_{Bt}$ are the total number of species A and B in demes, respectively. When $k_{A,B} = 0$, the above direction selection probabilities will reduce to those for selecting demes with the most vacancies, which is within one's general expectations.

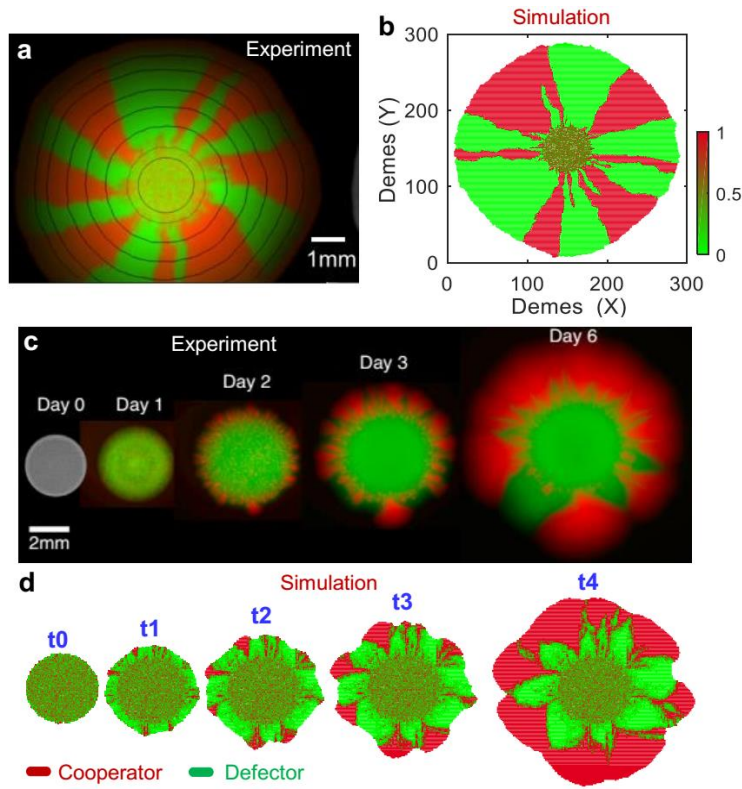

**Figure 3. Evolutionary games in biological range expansion.** (**a**), (**b**) Genetic demixing in range expansion of microorganisms: (**a**) Experiment on *E. coli* strains fluorescently labeled into the red and green [24]. (**b**) 2D on-lattice simulation (spatial distribution of frequency of the red strain): Maximum population size of each deme N = 30, migration rate m = 0.1 h$^{-1}$ and growth rate of strains w = 1 h$^{-1}$. These parameters give deme size a ≈ 6 um. (**c**),(**d**) Evolution of cooperation in Prisoner's dilemma: (**c**) Experiment: This microbial cooperator-defector system is constructed on *yeast* by using the synthetic biology method [18]. (**d**) 2D on-lattice simulation: Growth rates in the payoff matrix are respectively 0.3 h$^{-1}$, 0, 0.5 h$^{-1}$ and 0.1 h$^{-1}$, N = 20, migration rate

m = 0.1 h$^{-1}$. Cooperators are in red and defectors in green.

## 3. Model validation

Fig. 3a shows two *E. coli* strains, which defer only in expressing the green or red fluorescent protein, were well mixed initially and then inoculated to the solid agar. After several days, the colony developed into well-defined sector-like regions. This experiment [24] and the corresponding stepping stone models [20] showed that the formation of genetic de-mixing (i.e., spatial segregation between species) is determined by genetic drift resulting from the small number of reproductive individuals at the expanding frontier. Shown in Fig. 3b is our simulation result based on the 2D on-lattice simulation. The remarkable similarity with the experiment shows the credibility of our models.

Fig. 3c is the experimental results for the evolution of cooperation in Prisoner's dilemma [18]. Our simulation results in Fig. 3d show the similar evolutionary process, presenting again the high precision of our models and simulation techniques.

## 4. Results and discussion

Biological range expansion has been shown to be able to promote the evolution of cooperation in both numerical and experimental studies [16-19]. Spatial segregation produced in this process plays a key role [16]. We start with the discussion of this process here because a clear understanding still remains elusive.

### 4.1 The evolution of cooperation in linear range expansion

We show in Fig. 4 the evolutionary dynamics of cooperation in the Prisoner's dilemma game, which clearly reflects how cooperation is selected. Due to competition during range expansion, both cooperators and defectors can dominate in some local regions and generate spatial segregation from one another. As a natural protective shield arising from the proliferation of individuals (Fig. 5a), spatial segregation is able to further reduce the chance for cooperators to play games with defectors, resulting in cooperators having a higher proliferation ability and subsequent faster expansion speed (Fig. 5b). Consequently, defectors are gradually left behind and finally fail in the game. This competition process is vividly presented as the evolution in the number of individuals at the expanding frontier (Fig. 5c). In the competition with defectors, cooperators will first go through an adaptation phase. During this period, they suffer from fitness damage and have a reducing abundance within populations. After about 30 generations, however, cooperators develop a higher fitness, increase their abundance, and finally win the game. The terminal selection of cooperation in this game is influenced by the founder effect (such as the founder cell density and the founder frequency of cooperators) and the migration ability of individuals. Cooperation is more readily favored under lower cell density (Fig. 5d), lower migration rate (Fig. 5e) and higher initial frequency [18], since these situations will reduce the probability for cooperators to escape from defectors to establish local spatial segregation.

To identify which one between c and d is more advantageous for evolution of cooperation, which is crucial to disclose the fundamental cause for the evolution of cooperation, we further explored the selection outcomes of cooperation for c + d = const. (Figs. 5f, g). Simulations show an evolution law that the probability that cooperation can evolve in Prisoner's dilemma games linearly increases with the reduction in growth rate of defectors, although the total benefits defectors can get from cooperators and themselves are constant. This law illustrates that the maintenance of cooperation is achieved more readily through enhancing the self-proliferation advantage of cooperation relative to defection a-d, rather than reducing the benefits that defectors acquire from cooperators.

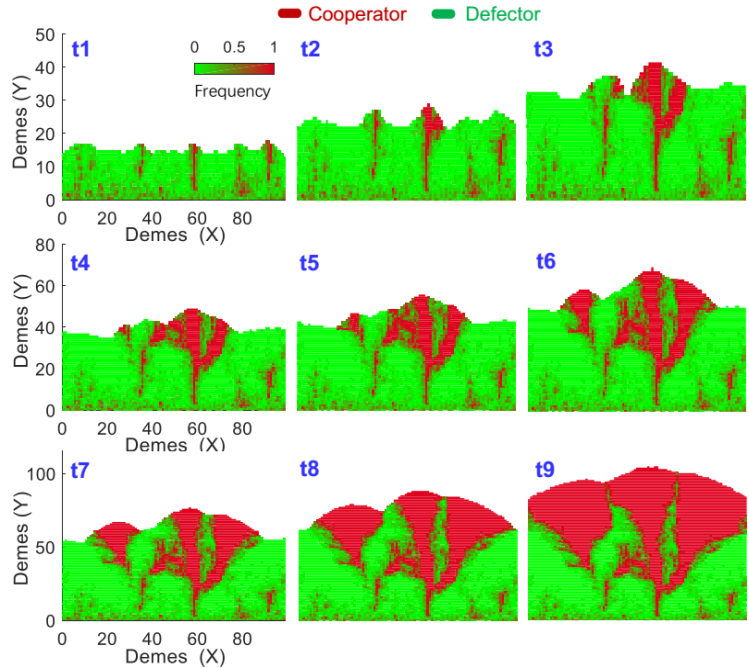

**Figure 4. 2D on-lattice simulation: evolutionary dynamics of cooperation in Prisoner's dilemma (linear range expansion).** Cooperation is gradually selected due to the spatial segregation generated by the higher self-proliferation of cooperators. N = 20 and m = 0.1 h$^{-1}$. Elements a, b, c and d (i.e. growth rates) in the payoff matrix are respectively 0.3 h$^{-1}$, 0, 0.5 h$^{-1}$ and 0.1 h$^{-1}$.

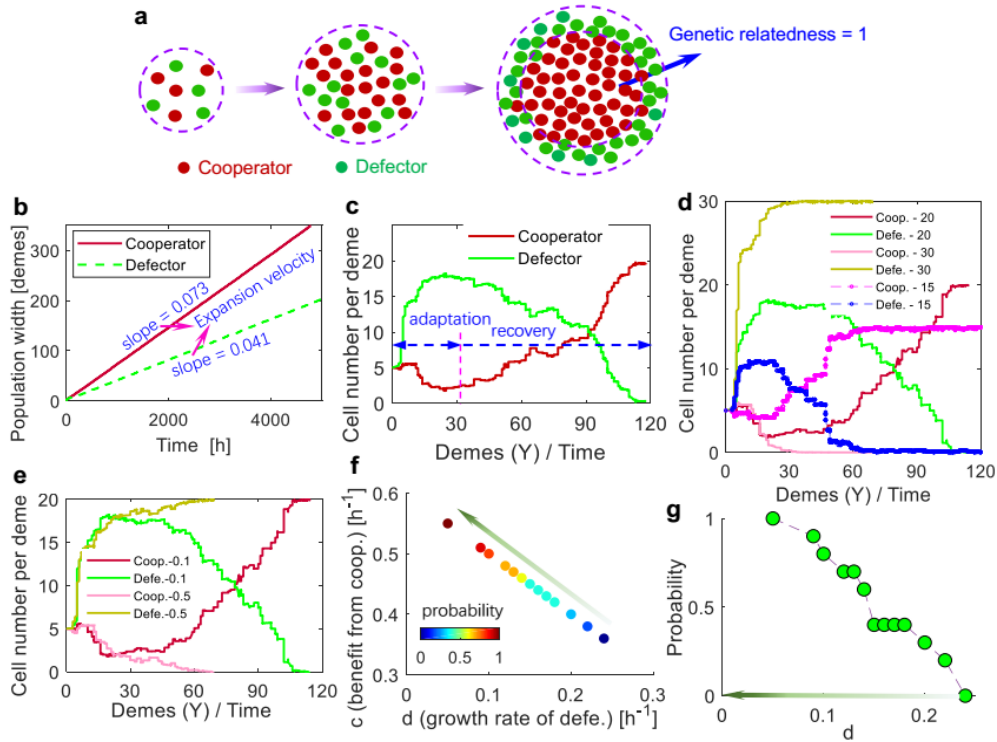

**Figure 5. Prisoner's dilemma games between cooperator and defector strains.** (**a**) Illustration of spatial segregation created by the proliferation of individuals. (**b**) Evolution of the population length in y direction when cooperators and defectors are grown individually: Growth rates of cooperators and defectors are 0.3 h$^{-1}$ and 0.1 h$^{-1}$ (N = 20 and m = 0.1 h$^{-1}$), respectively. (**c**) Evolutionary dynamics of cell number at the root of expanding frontier: Growth rates in the payoff matrix are respectively 0.3 h$^{-1}$, 0, 0.5 h$^{-1}$ and 0.1 h$^{-1}$. (**d**), (**e**) Evolutionary dynamics of the Prisoner's dilemma games under different (**d**) cell density N (m =

0.1 h$^{-1}$) and (**e**) migration rates m, h$^{-1}$ (N = 20): Initial cell number of both cooperators and defectors is controlled as 5 per deme. (**f**), (**g**) Evolution probability of cooperation (**f**) *vs*. c + d = const. and (**g**) *vs*. d in Prisoner's dilemma: Each point is tested for 10 rounds (N = 20 and m = 0.1 h$^{-1}$).

The inherent self-reproduction advantage of cooperators, which establishes spatial segregation (geographical isolation) from defectors, plays a significant role in promoting the evolution of cooperation. To enhance it is a more effective channel than to weaken the benefit capture of defectors from cooperators. Such a mechanism is simple as it does not require the organism to have the ability of implementing complex protective mechanisms like kin selection. We need to stress that the fundamental evolutionary advantage of cooperation in range expansion lies in the higher self-proliferation ability of cooperators themselves, resulting from the individual cooperative behavior with which cooperators are able to colonize virgin space in advance. This result is important as it shows that conducting cooperative behavior without retaliation (e.g. 'tit for tat' strategy [31,32]) is still favored.

### 4.2 Chemotaxis effect on the evolution of populations

Chemotaxis behavior, the directed motion guided by beneficial chemical cues, is a fitness-enhancing mechanism exhibited by many organisms, from bacteria [33] to insects [34] to animals [35]. Chemotaxis is a product of natural selection and, intuitively, one might expect chemotaxis to promote the evolution of populations. However, we identified a counterintuitive observation on the competition outcome between cooperators and defectors during radial range expansion (Fig. 6).

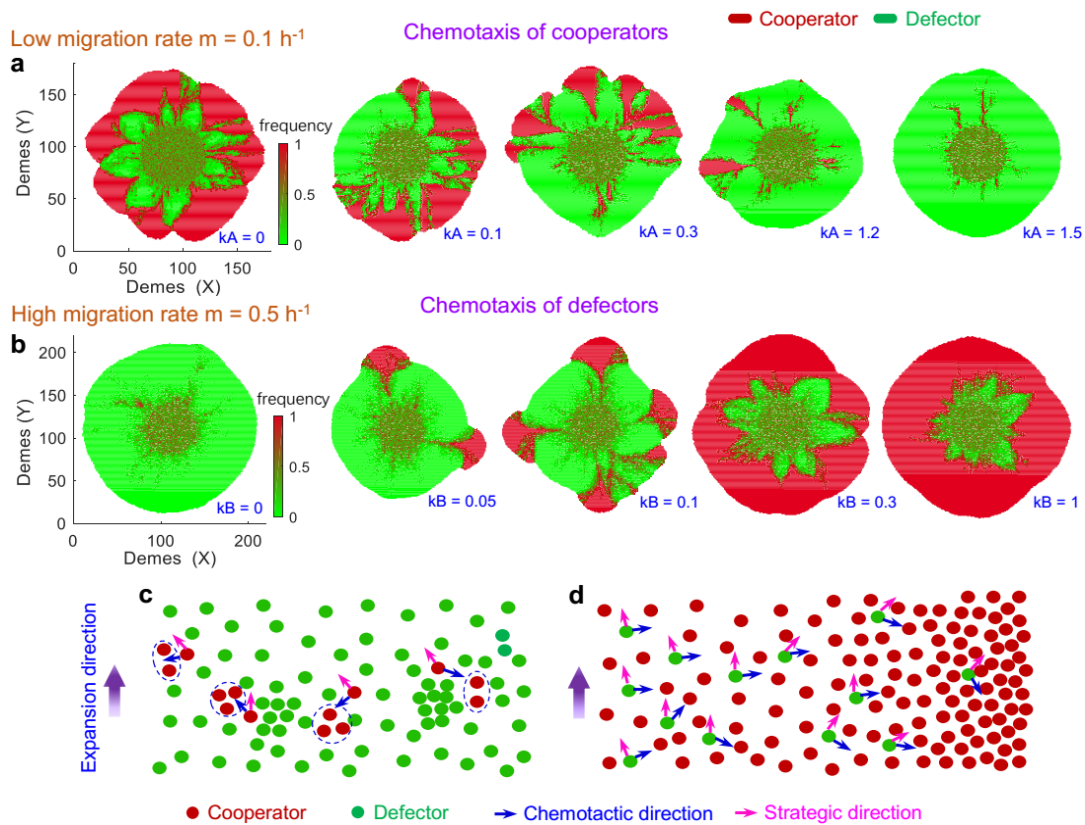

**Figure 6. 2D on-lattice simulation of the radial range expansion (*E. coli* strains): chemotaxis effect on the evolution of cooperation in Prisoner's dilemma.** A deme represents a small population, located on a spatial lattice. (**a**) Evolutionary games with the incorporation of chemotaxis of single cooperators and lower migration rate of individuals m = 0.1 h$^{-1}$: With an increase

in chemotaxis strength, evolutionary outcome transitions from cooperation to defection. (**b**) Evolutionary games with the incorporation of chemotaxis of single defectors and higher migration rate of individuals m = 0.5 h$^{-1}$: With the increase in chemotaxis strength, cooperation is increasingly favored. (**c**), (**d**) Illustration of the discrepancy between individual and population-level profits at the expanding frontier: (**c**) Chemotaxis of cooperators: Cooperators might move into a deme (location) where no defectors exist, but several cooperators might exist, which reduces the probability of them colonizing virgin space in advance. Cooperation may not be favored by competition in such cases. (**d**) Chemotaxis of defectors: Defectors are more likely to move into the demes with more cooperators, which will leave aside a minority of cooperators which are at low density to survive and reproduce and lead defectors to be in unfavorable situation. Parameters $k_A$ and $k_B$ represent the chemotaxis strength, the population size of each deme $N$ = 20. Growth rates in the payoff matrix are respectively 0.3 h$^{-1}$, 0, 0.5 h$^{-1}$ and 0.1 h$^{-1}$.

Within the Prisoner's dilemma, defectors would cause damage to cooperators through social interactions between them. We wanted to explore what would happen if cooperators or defectors have an extra ability to evade their opponents (defectors) or chase after benefits (cooperators) through chemotactic movement. In our simulations, we assume defectors can release some chemical, with which cooperators tend to move away from defectors. At the same time, defectors can use the chemicals released by cooperators to track cooperators for more benefits (see Section 2.2). We observe that chemotaxis of individuals impedes, rather than promotes, their competition with opponents at the population level (Figs. 6a and 6b).

As in Fig. 6a, cooperation can be selected during range expansion if the spatial segregation can be established in a timely manner [18] (e.g. under the low migration rates, which means less frequent encounters with defectors). When the chemotaxis of cooperators is incorporated, with an increase in chemotaxis strength evolution outcome is observed to transition from cooperation into defection. Similarly, in Fig. 6b, defection is selected under a higher migration rate of individuals, when no chemotaxis is involved. With the incorporation of chemotaxis of defectors and the increase in chemotaxis strength, natural selection increasingly favors cooperation.

The phenomenon above is not very hard to understand since natural selection is a short-sighted process [15]. Without chemotaxis, both cooperators and defectors tend to colonize virgin habitat in the expansion direction of their populations (move in the strategic direction) (Figs. 6c, d). In such situations, individual profits are nearly consistent with their populations. With the involvement of chemotaxis, however, each cooperator and defector have their own judgment on personal profits based on directed motion, which will reduce the fit between the interests of individuals and populations. As shown in Fig. 6c, cooperators tend to evade defectors through the chemotactic movement. In such a case, moving to locations without defectors is the most beneficial to cooperators. However, such a situation will cause cooperators to have a probability to move into a deme (location) where no defectors exist, but several cooperators have existed. Compared with singly migrating into the virgin space without both cooperators and defectors in the range expansion direction when no chemotaxis is involved, the involvement of the chemotactic movement reduces the probability of cooperators colonizing virgin space in advance. In considering the Prisoner's dilemma, however, the evolutionary advantage for cooperation is to colonize virgin space in the range expansion direction (Fig. 4). Cooperation may not be favored by competition in such cases. In Fig. 6d, defectors are more likely to move into the demes with more cooperators through the chemotactic movement, which will leave aside cooperators at the low density to survive and reproduce. Defectors will be always directed by cooperators in this case and are therefore in unfavorable position in evolution.

Our result demonstrates that there is a conflict between individual and population-level interests (Figs. 6c and 6d), and in the competition between populations, an individual's evolutionary advantage might not be that of a population. Instead, it may even become deleterious to a population's evolution, and the short-sighted selfish behaviors of individuals may not be that favored as we expect. This situation sets a higher requirement for the evolution of populations.

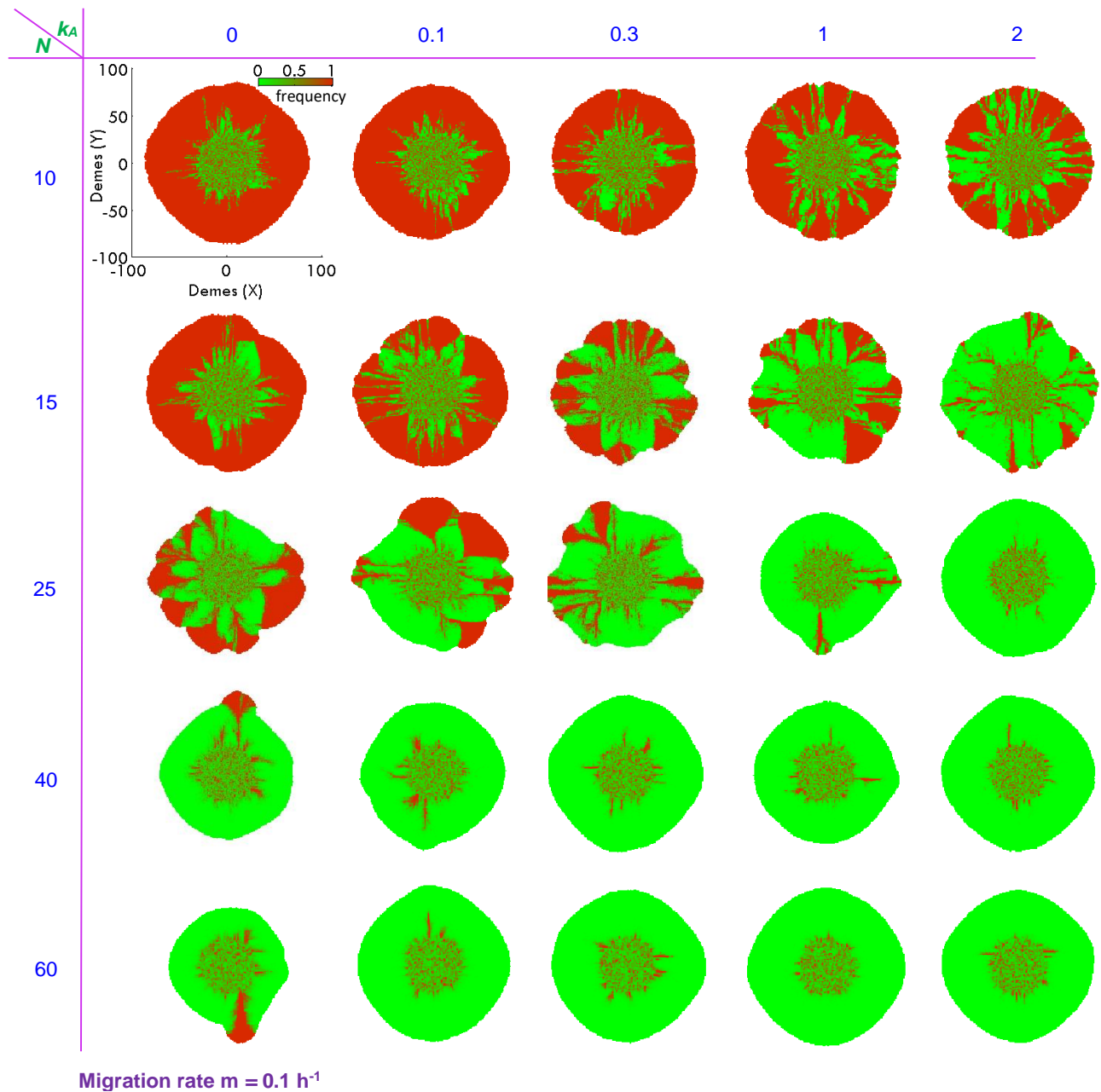

**Figure 7. 2D on-lattice simulation: Evolutionary games between cooperators and defectors in prisoner's dilemma for varying cell density $N$ and chemotaxis strength $k_A$.** Only the chemotaxis of cooperators is included. Growth rates in the payoff matrix are respectively 0.3 h$^{-1}$, 0, 0.5 h$^{-1}$ and 0.1 h$^{-1}$. Migration rate of individuals m = 0.1 h$^{-1}$.

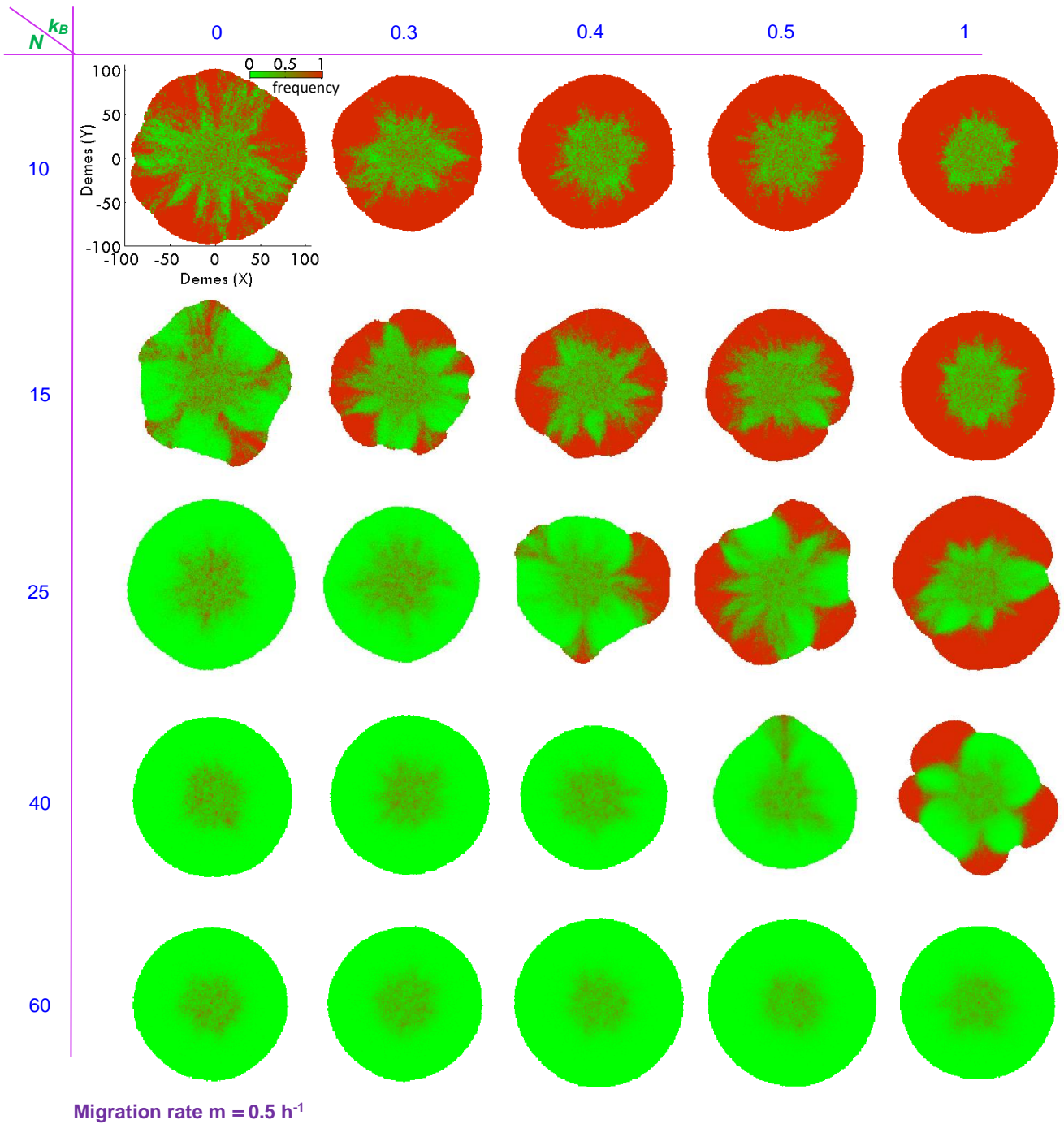

**Figure 8. 2D on-lattice simulation: Evolutionary games between cooperators and defectors in prisoner's dilemma for varying cell density *N* and chemotaxis strength $k_B$.** Only the chemotaxis of defectors is included. Growth rates in the payoff matrix are respectively 0.3 h$^{-1}$, 0, 0.5 h$^{-1}$ and 0.1 h$^{-1}$. Migration rate of individuals m = 0.5 h$^{-1}$.

We further check the effect of cell density *N*. Phase diagrams (Figs. 7 and 8) show again that higher cell density will improve the difficulty for cooperators to outcompete defectors, as analyzed in Fig. 5d. Such an effect causes lower chemotaxis strength for cooperators to transition the evolutionary outcome from cooperation to defection (Fig. 7), but higher chemotaxis strength for defectors to transition the evolutionary outcome from defection to cooperation (Fig. 8). As in Fig. 7 where only the chemotactic movement of cooperators is involved, defectors can outcompete the game only if $k_A > 2$ when *N* = 15, but this requirement for defectors is reduced to $k_A > 0.3$ when $N = 25$. For the case that only the chemotaxis of defectors exists (Fig. 8), the requirement for cooperators to defeat defectors is improved from $k_B > 0.3$ for $N = 15$ to $k_B > 1$ for $N = 40$.

## 5. Summary

Cooperation is responsible for biological organizations at all levels. Explaining cooperation is not straightforward since cooperation is usually believed to have a smaller competitive advantage than selfish behaviors like defection. Although many theories have been proposed to explain cooperation, cooperation's evolution is still an open question. We stress here that the inherent self-reproduction advantage of cooperators relative to defectors plays a fundamentally important role in promoting the evolution of cooperation in range expansion. This advantage leads cooperators to rapidly colonize virgin space in advance and establish the spatial segregation from defectors, which further promotes the evolution of cooperation. We show that enhancing this advantage is a more effective channel than dampening the benefit acquisition of defectors from cooperators. We also identify an interesting ecological mechanism that in games between populations, the fitness-enhancing chemotactic movement of individuals impedes the evolution of the population. This suggests that an individual's evolutionary advantage (e.g. the chemotaxis) might not be that of a population, and it can even be deleterious to the evolution of the population, as there is a conflict between an individual's and the group's fitness. These results give deeper insights into the evolution of cooperation.

**Acknowledgements:** This work is supported by Zhejiang University, National Natural Science Foundation of China and the research builds on research on the calibration of gene expression experiments funded by the UK BBSRC (BB/E001742/1). We are grateful to comments from Antonio Marco on an earlier version of the manuscript.

**Author contributions**: X.W. conceived this research, developed models, implemented theoretical calculations and data analysis, and wrote the paper. A.H. evaluated the article, provided revision suggestions and contributed to the writing. All authors participated in discussion.

**Conflict of interest**: The authors declare no conflict of interest.

**Data and materials availability**: All calculation data are available from X.W. upon reasonable request.